\documentclass{ws-procs10x7}
\usepackage{balance}

\newcolumntype{d}[1]{D{.}{.}{#1}}

\def\Journal#1#2#3#4{{\it #1} {\bf #2}, #3 (#4)}

\newcommand{\ie}{{i.e.}}
\newcommand{\eg}{{e.g.}}

\makeindex
\begin{document}

\title{Muon Cooling and Future Muon Facilities\footnotemark}

\author{D. M. Kaplan$^\dagger$}

\address{Physics Division, Illinois Institute of Technology, Chicago, Illinois 60616, USA\\$^\dagger$E-mail: kaplan@iit.edu}

\twocolumn[\maketitle\abstract{Muon colliders and neutrino factories are attractive options for achieving the highest lepton-antilepton collision energies and the most precise measurements of the parameters of the neutrino mixing matrix. The performance and cost of these future facilities depends sensitively on how well a beam of muons can be cooled. The recent progress of muon-cooling prototype tests and design studies  nourishes the hope that such facilities can be built during the next decade.}
]
\section{Introduction}

\footnotetext{* To appear in Proceedings of the XXXIII International Conference on High Energy Physics, Moscow, Russia, July 26 -- August 2, 2006.}

\begin{figure}[tb]
\vspace{0.5in}
\includegraphics[bb=0 170 590 570,width=\linewidth]{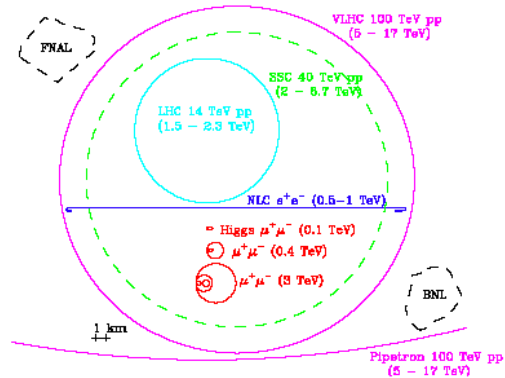}
\caption{Sizes of various proposed  colliders compared with FNAL
and BNL sites. A muon collider with $\sqrt{s}>3$\,TeV fits on existing
sites.}\label{fig:sizes}
\end{figure}

The muon offers important advantages over the electron for 
use in a high-energy collider: 
\begin{enumerate}
\item The $1/m^2$ suppression of radiative processes enables the use of storage rings and
recirculating accelerators, reducing the size (Fig.~\ref{fig:sizes}) and cost of the complex.
\item In the Standard Model and many  extensions, the muon/electron cross-section ratio for $s$-channel annihilation to Higgs bosons is $({m_\mu}/{m_e})^2=4.3\times10^4$, giving the muon collider a unique window on electroweak symmetry breaking.\cite{Bargeretal,strong-dynamics}
\item ``Beamstrahlung" interactions, which limit 
$e^+e^-$-collider luminosity as energy  increases,\cite{Palmer-Gallardo} are negligible for muons.
\end{enumerate}
Moreover, a  muon storage ring is  an ideal source for long-baseline neutrino-oscillation experiments: via $\mu^-\to e^-\nu_\mu{\overline \nu_e}$ and $\mu^+\to e^+{\overline \nu_\mu}\nu_e$, it can provide collimated, high-energy neutrino 
beams with well-understood composition and properties.\cite{Geer-ultimate}  The very clean identification of final-state muons in far detectors  
enables low-background appearance measurements using $\nu_e$ and $\overline \nu_e$ beams. The separation of oscillated from non-oscillated events requires only that the detector be magnetized so as to distinguish $\mu^+$ (the oscillated events if $\mu^-$ are stored in the ring) from $\mu^-$ (the oscillated events if $\mu^+$ are stored).

\begin{figure}[t]
\includegraphics[bb=0 100 585 690,width=\linewidth]{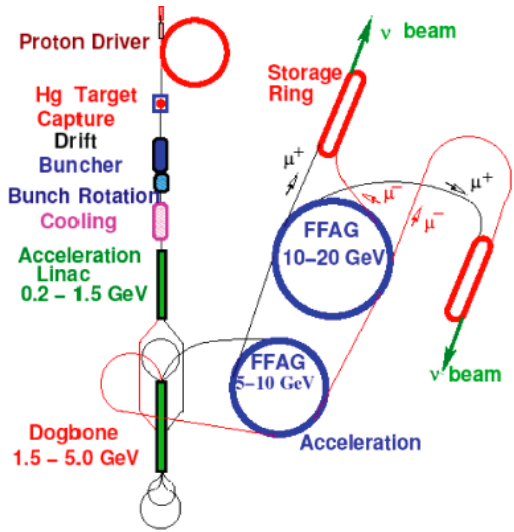}
\caption{Sketch of a recent neutrino-factory design:\protect\cite{ISS} pions created by beam from high-intensity ``proton driver" are captured and decay in a focusing channel; decay muons undergo phase-space manipulations, including transverse ionization cooling; are accelerated in a linac, a ``dogbone" recirculating linac (RLA), and two fixed-field alternating-gradient (FFAG) accelerators; and are stored in  two racetrack-shaped decay rings whose long straight sections each form oppositely directed neutrino and antineutrino beams aimed at near and far detectors.} 
\label{fig:ISS} 
\end{figure}

\begin{figure}[b]
\includegraphics[bb=20 0 570 770,width=\linewidth]{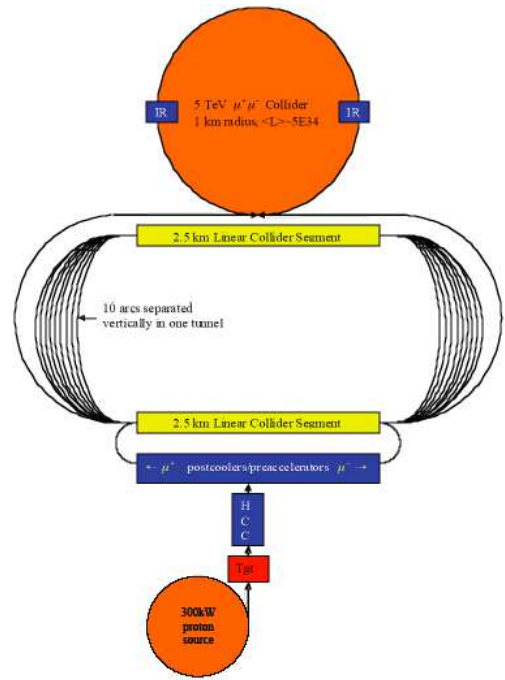}
\caption{Sketch of a Muons, Inc.\ muon collider concept with $\sqrt{s}=5$\,TeV: a helical cooling channel\protect\cite{recent}  (HCC) cools the $\mu^+$ and $\mu^-$ beams in all six dimensions sufficiently that they can then be accelerated in an RLA based on ILC-style rf-cavity modules.}\label{fig:mumu}
\end{figure}

These advantages come with
clear disadvantages: the short muon lifetime 
 and large beam size  
require development of new, rapid beam manipulation and acceleration techniques if intense muon beams are to be accelerated, stored, or collided. Stored-muon ``neutrino factories" (Fig.~\ref{fig:ISS}) and colliders (Fig.~\ref{fig:mumu}) benefit from muon-beam cooling,\cite{benefit} which allows smaller-aperture (hence less costly) accelerators 
and higher luminosity. 

\section{Muon Cooling}
Standard (electron, stochastic, and laser) beam-cooling methods are far too slow to be effective within the 2.2\,$\mu$s muon lifetime. However,  the muon's penetrating character enables rapid muon cooling via {\em ionization}.\cite{ionization-cooling,more-on-cooling} An ionization-cooling channel comprises energy absorbers and radio-frequency (rf) accelerating cavities placed within a  focusing magnetic lattice. In the absorbers the muons lose energy by ionization; the rf cavities restore energy only along the beam axis. In this way, the (initially highly divergent) muon beam can be made more parallel.

Cooling is best understood 
in terms of normalized beam emittance $\epsilon_n$, the volume of a beam in phase space, which is a constant of the motion both in linear beam transport and during acceleration. Cooling is the process of reducing a beam's normalized emittance. 
In a medium, normalized transverse emittance depends on path length $s$ as\cite{Ankenbrandt,Neuffer-yellow}
\begin{eqnarray*} 
\frac{d\epsilon_n}{ds}\approx
-\frac{1}{\beta^2} \left\langle\!\frac{dE_{\mu}}{ds}\!\!\right\rangle\frac{\epsilon_n}{E_{\mu}}
 +
\frac{1}{\beta^3} \frac{\beta_\perp
(0.014)^2}{2E_{\mu}m_{\mu}L_R},\ 
(1)
\label{eq:cool} 
\end{eqnarray*}  
where $\beta$ is the muon velocity in units of $c$,  $E_\mu$ the muon energy  in GeV, 
$m_\mu$
its mass in GeV/$c^2$, $\beta_\perp$ the lattice betatron function, and $L_R$ the radiation length of the medium. 
A portion of this cooling effect can be transferred to the longitudinal phase plane by  
placing suitably shaped absorbers  
in dispersive regions of the lattice (``emittance exchange")\cite{more-on-cooling,Ankenbrandt,Neuffer-yellow} or using path-length-dependent energy loss within a homogeneous absorber.\cite{homo} (Longitudinal ionization
cooling {\em per se} is impractical due to energy-loss straggling.\cite{Neuffer-yellow}) 

\label{sec:matl}
The terms of Eq.~\ref{eq:cool} 
represent muon cooling by  energy loss and heating by multiple Coulomb scattering. Setting the two terms equal gives the equilibrium
emittance $\epsilon_{n,eq}$, at which the cooling rate is zero and beyond which a given lattice cannot cool. Since the heating term scales with $\beta_\perp$, a low $\epsilon_{n,eq}$ requires
low  $\beta_\perp$ (\ie, high focusing strength) at the absorbers. Most design studies have used superconducting solenoids, which can give 
$\beta_\perp\sim10$\,cm, as the focusing element of choice. Concerning $L_R$, low-$Z$
absorber media are  
favored, the best being hydrogen (approximately twice as effective for cooling as helium, the next best material\cite{Kaplan-COOL03}). 

\begin{figure}[bt]
\includegraphics[bb=20 200 600 568,clip,width=\linewidth]{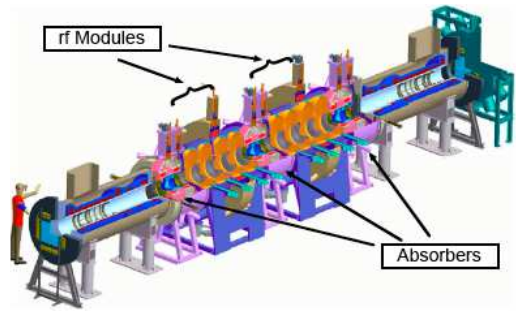}
\caption{Three-dimensional cutaway rendering of MICE apparatus (see text): muons entering at lower left are measured by time-of-flight (TOF) and Cherenkov counters and a solenoidal tracking spectrometer; then, in cooling section, alternately slowed  in LH$_2$ absorbers and reaccelerated by rf cavities, while focused by a lattice of superconducting solenoids; then remeasured by a second solenoidal tracking spectrometer and their muon identity confirmed by TOF detectors and a calorimeter.
}
\label{fig:MICE}
\end{figure}

It is the absorbers that cool the
beam, but for typical ``real-estate" accelerating gradients ($\approx$\,10\,MeV/m, to be compared with $\langle
dE_\mu/ds\rangle\approx30$ MeV/m for liquid hydrogen\cite{PDG}), the rf cavities dominate the
length of the cooling channel (see {\eg}\ Fig.~\ref{fig:MICE}). 
Ideally, the acceleration should exceed the minimum required for energy replacement, allowing
``off-crest" operation. This gives continual rebunching, so that 
a beam with large momentum
spread remains captured in the rf bucket. The achievable rf gradient
thus determines how much cooling is practical before an appreciable fraction of the muons have
decayed or drifted out of the bucket. High-gradient rf cavities (normal-conducting due to the magnetic field in which they must operate) for muon cooling are under development,\cite{RF} as is an alternative cooling approach: cavities pressurized with hydrogen gas, thus combining energy absorption and reacceleration.\cite{Hanlet} In the first cooling stages the large size of the uncooled beam requires relatively low rf frequency. Goals are $\stackrel{>}{_\sim}$\,15\,MeV/m at $\approx$\,201\,MHz in $\approx$\,2\,T fields.

In the cooling term of Eq.~\ref{eq:cool}, 
the percentage decrease in normalized emittance is proportional to the percentage energy loss, thus (approximating $\beta\approx1$) cooling in one transverse dimension by a factor 1/$e$ requires $\sim$\,100\% energy loss and replacement.  Despite the relativistic increase of muon
lifetime with energy, ionization cooling favors low beam momentum
 because of the increase of
$dE/ds$ for momenta below the ionization minimum,\cite{PDG} the greater ease of beam focusing, and the lower accelerating voltage required.  Most muon-cooling designs 
have used momenta in the range 150$-$400\,MeV/$c$. This is also the momentum range in which
the pion-production cross section from thick targets tends to peak and is thus optimal for muon
production as well as  cooling. The cooling channel of Fig.~\ref{fig:MICE} is optimized for a
mean muon momentum of 200\,MeV$/c$.  

\section{Towards a Muon Collider}

Cooling lattices using longitudinal--transverse emittance exchange to cool simultaneously in all six dimensions  are receiving increasing attention,\cite{Palmer-ring,recent}  from both the Neutrino Factory and Muon Collider Collaboration\cite{NFMCC} and Muons, Inc.\cite{Muonsinc} These are essential to a high-luminosity muon collider and may  enable  higher-performance or lower-cost neutrino factories. 
As Fig.~\ref{fig:mumu} suggests, muon colliders offer the prospect of much higher collision energies than are feasible with electrons; they thus provide a potential next step beyond the ILC.

\section{Technology Demonstrations}

The R\&D on muon cooling\cite{Alsharoa} has identified a number of technologies crucial to future muon facilities, each of which has a demonstration experiment proposed or in progress:
\begin{enumerate}

\item The MERIT (Mercury Intense Target) experiment, approved at CERN and under construction for operation in 2007; the goal is to show feasibility of a mercury-jet target for a 4\,MW proton beam with solenoidal pion capture.\cite{MERIT}

\item MICE (the Muon Ionization Cooling Experiment, see Fig.~\ref{fig:MICE}), approved at Rutherford Appleton Laboratory and under construction, aiming to verify the feasibility and performance of transverse ionization cooling by  2010.\cite{MICE}

\item EMMA (Electron Model of Muon Accelerator), proposal to build and operate at Daresbury Laboratory a model ``non-scaling" FFAG accelerator.\cite{EMMA}

\item MANX (Muon collider And Neutrino factory eXperiment), LoI to build and test a helical cooling channel 
segment.\cite{Muonsinc}

\end{enumerate}
Experimental results may soon strengthen the physics case for a muon facility. With the key techniques established by $\approx$2010, a facility could then be built in the ensuing decade.

\section{Acknowledgments}
This work was supported  by the US Dept.\ of Energy and the National Science Foundation.


\begin{thebibliography}{99}
\bibitem{benefit}
E.g., cooling is cost-effective and worth a factor $\approx$1.7 in $\nu$ intensity in the ISS design.\protect\cite{ISS}
\end{thebibliography}
\end{document}